# Interplay between unconventional superconductivity and heavy-fermion quantum criticality: $CeCu_2Si_2$ versus $YbRh_2Si_2$

M. Smidman[1], O. Stockert[2], J. Arndt[2], G. M. Pang[1], L. Jiao[1,2], H. Q. Yuan[1], H. A. Vieyra[2], S. Kitagawa[3], K. Ishida[3], K. Fujiwara[4,5], T. C. Kobayashi[5], E. Schuberth[6,2], M. Tippmann[6], L. Steinke[7,6,2], S. Lausberg[2], A. Steppke[2], M. Brando[2], H. Pfau[8,2], U. Stockert[2], P. Sun[9,2], S. Friedemann[10,2], S. Wirth[2], C. Krellner[11,2], S. Kirchner[12], E. M. Nica[13,14], R. Yu[15,14], Q. Si[14] and F. Steglich[2,1,9]

[1]Center for Correlated Matter, Zhejiang University, Hangzhou, Zhejiang 310058, China

[2]Max Planck Institute for Chemical Physics of Solids, 01187 Dresden, Germany

[3]Department of Physics, Kyoto University, Kyoto 606-8502, Japan

[4]Department of Materials Science, Shimane University, Matsue 690-8504, Japan

[5]Graduate School of Natural Science and Technology, Okayama University, Okayama 700-8530, Japan

[6]Walter Meissner Institute for Low Temperature Research, Bavarian Academy of Sciences, 85748 Garching, Germany

[7]Department of Physics and Astronomy, Texas A&M University, College Station, Texas 77843, USA

[8]Stanford Institute for Materials and Energy Sciences, SLAC National Accelerator Laboratory, Menlo Park, California 94025, USA

[9]Institute of Physics, Chinese Academy of Sciences, Beijing 100190, China

[10]School of Physics, H H Wills Physics Laboratory, University of Bristol, Bristol BS8 1TL, UK

[11]Physics Institute, University of Frankfurt, 60438 Frankfurt, Germany

[12]Zhejiang Institute of Modern Physics, Zhejiang University, Hangzhou, Zhejiang 310027, China

[13]Department of Physics and Astronomy and Quantum Materials Institute, University of British Columbia, Vancouver V6T 1Z1, Canada

[14]Department of Physics and Astronomy, Rice University, Houston, Texas 77005, USA

[15]Department of Physics, Renmin University of China, Beijing 100872, China

Corresponding authors: M. Smidman, email: <u>msmidman@zju.edu.cn</u>; F. Steglich, email: <u>frank.steglich@cpfs.mpg.de</u>

**Abstract**

In this paper the low-temperature properties of two isostructural canonical heavy-fermion compounds are contrasted with regards to the interplay between antiferromagnetic (AF) quantum criticality and superconductivity. For $CeCu_2Si_2$, fully-gapped *d*-wave superconductivity forms in the vicinity of an *itinerant* three-dimensional heavy-fermion spin-density-wave (SDW) quantum critical point (QCP). Inelastic neutron scattering results highlight that both quantum critical SDW fluctuations as well as Mott-type fluctuations of local magnetic moments contribute to the formation of Cooper pairs in $CeCu_2Si_2$. In $YbRh_2Si_2$, superconductivity appears to be suppressed at $T \gtrsim 10$ mK by AF order ($T_N$ = 70 mK). Ultra-low temperature measurements reveal a hybrid order between nuclear and 4*f*-electronic spins, which is dominated by the Yb-derived nuclear spins, to develop at $T_A$ slightly above 2 mK. The hybrid order turns out to strongly compete with the primary 4*f*-electronic order and to push the material towards its QCP. Apparently, this paves the way for heavy-fermion superconductivity to form at $T_c$ = 2 mK. Like the pressure - induced QCP in $CeRhIn_5$, the magnetic field - induced one in $YbRh_2Si_2$ is of the *local* Kondo-destroying variety which corresponds to a Mott-type transition at zero temperature. Therefore, these materials form the link between the large family of about fifty low-*T* unconventional heavy - fermion superconductors and other families of unconventional superconductors with higher $T_c$s, notably the doped Mott insulators of the cuprates, organic charge-transfer salts and some of the Fe-based superconductors. Our study suggests that heavy-fermion superconductivity near an AF QCP is a robust phenomenon.

## 1. Quantum criticality in antiferromagnetic heavy-fermion metals

Unconventional superconductivity, i.e., superconductivity which is not driven by lattice vibrations, frequently develops in strongly correlated metals on the brink of antiferromagnetic (AF) order [1, 2]. The continuous suppression of AF order by non-thermal control parameters, such as external/chemical pressure and magnetic field gives rise to a quantum critical point (QCP) which determines the physical properties in a wide range of parameters. Strong deviations from the predictions of Landau's Fermi-liquid theory [3], so-called non-Fermi-liquid (NFL) phenomena, are commonly observed in the normal metallic state out of which superconductivity develops. The interplay between quantum criticality and superconductivity in strongly correlated electron systems is a timely, controversial and much debated topic which has been studied over the last two decades, most intensively with AF heavy-fermion metals [4, 5]. These are intermetallic compounds of certain lanthanides, such as Ce and Yb, or actinides, such as U and Pu. The lanthanide-based heavy-fermion metals are model systems for the *Kondo lattice*, where at the QCP the on-site Kondo screening, characterized by $k_B T_K$, with $T_K$ being the Kondo temperature of the crystal-field (CF) - derived lowest-lying Kramers doublet of the localized 4*f*-shell, exactly cancels the intersite magnetic Ruderman Kittel Kasuya Yoshida (RKKY) interaction, characterized by $k_B T_{RKKY}$. So far, two different types of AF QCPs have been established for heavy-fermion metals. Some of them exhibit a "conventional" QCP, which means that in this scenario the AF order is of *itinerant* nature [6 - 8]. This kind of spin-density wave (SDW) order, with three-dimensional (3D) quantum critical fluctuations of the AF order parameter, is common to transition-metal compounds where *d*-electrons contribute to the conduction band. A heavy-fermion SDW QCP ensues once at low temperatures the heavy charge carriers, composite quasiparticles which are formed by the Kondo interaction and consist of *f*-electron as well as conduction-electron contributions, behave like *d*-electrons in that they exist on both sides of the instability. In contrast, inelastic neutron scattering (INS) results on $CeCu_{6-x}Au_x$ ($x = x_c = 0.1$) revealed a dynamical susceptibility with an unusual $\omega/T$ scaling and a fractional exponent, not only at the ordering wave vector of the AF phase that forms at $x > x_c$, but also in an extended part of the Brillouin zone [9]. This finding has motivated theorists to come up with proposals of an "unconventional", i.e., *local*, QCP [10-12]. Here, the disappearance of AF order at $T = 0$ concurs with the destruction of the lattice Kondo effect, i.e., the disintegration of the composite heavy quasiparticles. At such a Kondo-destroying QCP the Fermi-surface volume was predicted to change abruptly from "small" (conduction electrons only) to "large" (conduction electrons and *f*- electrons) [10, 11]. This prediction was verified by direct Fermi-surface measurements via de Haas - van Alphen oscillations under pressure for the local-moment antiferromagnet $CeRhIn_5$ [13]. The study showed that (i) the Fermi surface volume jumps at the pressure-induced AF QCP, $p_c \approx 2.4$ GPa, and (ii) the cyclotron mass increases significantly upon approaching $p_c$ from either side. Both observations are consistent with a local QCP. For this compound, a dome of heavy-fermion superconductivity exists at finite pressures, $p \geq 1$ GPa, and is almost centered around $p = p_c$ [14, 15]. In view of the fragile electronic structure near the Kondo-destroying QCP, the thermodynamic stability of the superconducting phase appears to be surprisingly strong: The maximum $T_c$ of pressurized $CeRhIn_5$ is 2.3 K, almost as large as the highest $T_c$ observed in a Ce-based heavy fermion compound (2.5 K for $CeAu_2Si_2$ [16]).

In this article, we wish to address superconductivity in two heavy-fermion compounds crystallizing in the same tetragonal $ThCr_2Si_2$ structure, each of which being a prototypical material for one of the two different QCP scenarios introduced above: $CeCu_2Si_2$ exhibiting a volume-dependent *itinerant* (3D-SDW) QCP [17] and $YbRh_2Si_2$ with a magnetic field-induced *local* QCP [18]. The former system shows the unique behavior of a fully-gapped *d* - wave superconductor [19], while heavy-fermion superconductivity in $YbRh_2Si_2$ emerges at ultra-low temperatures under the influence of nuclear AF order [20].

## 2. $CeCu_2Si_2$: Fully gapped *d*-wave superconductivity in the vicinity of a three-dimensional spin-density-wave quantum critical point

Heavy-fermion superconductivity was first discovered in the Kondo lattice system $CeCu_2Si_2$ with almost trivalent Ce ($T_K \simeq 20$ K) [21]. It has been considered an unconventional superconductor from early on: (i) The non-*f*-electron reference compound $LaCu_2Si_2$ does not superconduct (at $T \geq 20$ mK) [21], which implies that superconductivity in the Ce homologue should be ascribed to the periodic lattice of 100 % magnetic $Ce^{3+}$ ions. (ii) The reduced jump in the Sommerfeld coefficient of the electronic specific heat [$\gamma(T) = C(T)/T$] at the superconducting transition temperature, $\Delta C/\gamma_0 T_c$ [$\gamma_0 \simeq 1$ J/(K$^2$mol)], is of order unity, which implies that the Cooper pairs are formed by heavy-mass quasiparticles, i.e., slowly propagating Kondo singlets. As their Fermi velocity $v_F$* is only of the order of the velocity of sound, the electron-phonon interaction is not retarded, i.e., the direct Coulomb repulsion among the charge carriers cannot be avoided. (iii) Therefore, an alternative pairing mechanism must be at work which, in analogy to superfluidity in $^3$He [22], was early on assumed to be magnetic in origin [23 - 25]. (iv) Already a tiny amount of *nonmagnetic* impurities was found to fully suppress superconductivity in $CeCu_2Si_2$ [26], similar to the effect *magnetic* impurities have in a conventional (BCS) superconductor [27, 28]. Also, low-temperature irradiation of $CeCu_2Si_2$ thin films with fast ions has a strong pair-breaking effect [29]. Because of a substantial Pauli limiting effect inferred from measurements of the upper critical magnetic field $B_{c2}(T)$ [30, 31], the superconducting state turned out to be an even-parity (spin-singlet), presumably *d*-wave state, as discussed below. This is in contrast to (spin-triplet) *p*-wave superfluidity in $^3$He [22]. Refs. 30 and 31 report a huge slope of $B_{c2}$ at $T_c$ which confirms the conclusion from calorimetric studies [21] that in $CeCu_2Si_2$ Cooper pairs are built of heavy-mass charge carriers.

The first experimental indication of magnetically driven superconductivity was the observation of a sizable decrease of the magnetic neutron-scattering intensity below $T_c \simeq$ 0.5 K in the heavy-fermion superconductor $UPt_3$ which shows weak AF order below $\simeq$ 5 K [32]. Later, in the local-moment antiferromagnetically ordered ($T_N$ = 14.3 K) heavy-fermion superconductor $UPd_2Al_3$ a comparison between quasiparticle tunnelling [33] and INS results [34] revealed a Cooper pairing driven by "magnetic excitons" (at the ordering wave vector), which are dispersive crystal-field excitations that exist already in the paramagnetic state and become acoustic magnons below $T_N$.

$CeCu_2Si_2$ is located very close to an AF QCP which, owing to the $\gamma = \gamma_0 - bT^{1/2}$ dependence of the Sommerfeld coefficient and the $\rho = \rho_0 + aT^{3/2}$ dependence of the electrical resistivity in the low-temperature normal state, has been considered to be of the itinerant 3D-SDW type [35]. Here, $\rho_0$ and $\gamma_0$ are the residual, finite values as $T \to 0$. The latter one implies that the effective quasiparticle mass does not diverge at this QCP which, for this reason, is often termed "conventional". $CeCu_2Si_2$ can be tuned through its QCP by the application of only moderate hydrostatic pressure or by composition/chemical substitution. Superconductivity appears around the QCP. Hence, the AF order and the superconductivity in $CeCu_2Si_2$ are easily accessible experimentally. Within a narrow homogeneity range, this material shows different ground states depending on the actual composition: Si-rich samples exhibit AF order ('*A*-type' $CeCu_2Si_2$) while Cu excess leads to superconducting samples ('*S*-type') [36]. Nearly stoichiometric $CeCu_2Si_2$ displays AF order at elevated temperatures, followed by a first-order transition into superconductivity at lower temperatures where both phenomena exclude each other on a microscopic scale ('*A/S*-type' samples), see Ref. 37. Interestingly, $T_c$ in *S* - and *A/S* - type $CeCu_2Si_2$ samples turns out to be robust against variations of the residual resistivity by a factor of order four (see below) – $\rho_0$ being typically substantially large ($\geq$ 10 μΩcm). This insensitivity of SC against "irrelevant" disorder, as manifested by $\rho_0$, vis à vis the extreme pair - breaking capability of substitutional disorder [26] in $CeCu_2Si_2$ remains an interesting issue for future studies, see below.

The first neutron-scattering experiments on $CeCu_2Si_2$ focused on the AF order in *A*-type crystals. Single-crystal neutron diffraction revealed an incommensurate SDW nature of the antiferromagnetism. As displayed in the intensity maps in Fig. 1a, magnetic superstructure peaks were detected in the (*hhl*) scattering plane below a Néel temperature $T_N \approx$ 0.8 K, which agrees with that obtained from thermodynamic measurements [35]. Their positions yield an incommensurate AF structure with an ordering wave vector $\boldsymbol{Q}_{AF}$ = (0.215 0.215 0.53) with respect to a nuclear Bragg reflection, and an ordered magnetic moment of $\approx$ 0.1$\mu_B$ at $T$ = 0.05 K. $\boldsymbol{Q}_{AF}$, found at any symmetry-related position, is *identical* to the nesting wave vector which connects the parallel, flat parts of the dominating heavy-fermion pocket

(warped cylinders along the *c*-direction) [17]. Approaching the QCP, i.e., studying an *S*-type single crystal, static long-range antiferromagnetism appears to be fully suppressed, and only dynamical spin correlations remain (it is worth mentioning that small isolated magnetically ordered, phase separated, regions of 100 Å size exist in the *S* – type crystal studied, with a total volume fraction ≤ 10% [38]).

Upon investigating the (quantum-)critical spin fluctuations at $Q_{AF}$ at the critical magnetic field necessary to fully suppress superconductivity in *S* - type $CeCu_2Si_2$, $B = B_{c2}$ = 1.7 T, one finds $\omega/T^{3/2}$ scaling of the imaginary part of the dynamical spin susceptibility $\chi''(T)$ over wide ranges of energy transfer and temperature (see Fig. 1b). Together with the $T^{3/2}$ dependence of the inverse lifetime of the spin fluctuations at $Q_{AF}$ seen in Fig. 1d, this points to an (almost) critical slowing down of the magnetic response and is direct support for a QCP with 3D-SDW criticality [39]. While the normal-state magnetic response in *S*-type $CeCu_2Si_2$ is quasielastic, the response becomes inelastic on entering the superconducting state, and a spin gap opens with the accumulation of spectral weight just above the gap resulting in an inelastic peak in $\chi''(\hbar\omega)$, as displayed in Fig. 1c [38]. This spin resonance occurs at a value of ≈ 0.2 meV at the lowest measured temperature of 70 mK which corresponds to $\hbar\omega$ = 3.9 $k_BT_c$. This is about 20% smaller than $2\Delta(0) \simeq 5\ k_BT_c$, $\Delta(0)$ being the larger one of the two superconducting gaps at *T* = 0 discussed below.

The magnetic responses in the normal and superconducting states are not limited just to $Q_{AF}$, but the spin dynamics are overdamped paramagnons and exhibit a roughly linear dispersion around $Q_{AF}$ in momentum-energy space as indicated in Fig. 1e [38, 39]. The main difference between the normal and the superconducting states occurs very close to $Q_{AF}$, where a spin-excitation gap opens in the superconducting state below 0.2 meV. The knowledge of $\chi''(q, \omega)$ in the relevant part of the Brillouin zone allows for a calculation of the magnetic exchange energy. As a result, the gain in magnetic exchange energy upon entering the superconducting state is larger by a factor of about 20 than the superconducting condensation energy determined via heat-capacity measurements on the same sample [38]. For this reason, the almost critical spin fluctuations can be considered to be contributing substantially to the formation of unconventional superconductivity in $CeCu_2Si_2$.

The formation of a spin gap with a pronounced spin resonance in the INS intensity deeply within the superconducting gap of $CeCu_2Si_2$ yields information about the symmetry of the gap function. In a BCS superconductor, the quasiparticle contribution to the dynamical spin susceptibility in the one-band case [41, 42] has the following form (generalizations to the multi-band case are discussed by E. M. Nica et al. [43]):

$$\chi_0(q,\omega) \sim \sum_k \{1-[(\varepsilon_{k+q}\varepsilon_k + \Delta_{k+q}\Delta_k)/(E_{k+q}E_k)]\}[(f(E_{k+q}) + f(E_k) -1)]/[\omega - (E_{k+q} + E_k) + i0^+], \qquad (1)$$

where $\Delta_k$ is the gap function, $\varepsilon_k$ describes the normal-state dispersion and $E_k$ the corresponding Bogoliubov-de Gennes (BdG) quasiparticle dispersion, *f* is the Fermi-Dirac function and $q = Q_{AF}$. The first factor on the r.h.s. of Eq. 1 denotes the spin coherence factor. For a sign-changing gap function, $\Delta_{k+q}\Delta_k = -|\Delta_{k+q}||\Delta_k|$, and the coherence factor is nonzero. In this case, the magnetic exchange interactions between the BdG quasiparticles give rise to a sharp peak *below* twice the characteristic superconducting gap magnitude $2\Delta$ [44, 45]. For a sign-preserving gap function, $\Delta_{k+q}\Delta_k = |\Delta_{k+q}||\Delta_k|$, and the coherence factor is vanishingly small. The spin spectrum will not have a sharp peak. At most, it would display only some

broad enhancement of spectral weight *above* 2$\Delta$ [44, 45]. The above considerations can be equivalently given in terms of a simplified form of the time-reversal odd coherence factor entering Eq. 1, which is 1-cos[$\emptyset(\mathbf{q})$] with $\emptyset(\mathbf{q})$ being the phase difference between the gap values at two wavevectors on the Fermi surface, $\mathbf{q}$ and $\mathbf{q}+\mathbf{Q_{AF}}$, respectively [46, 47]. This factor vanishes for a sign-preserving gap function with $\emptyset(\mathbf{q})$ = 0, but is nonzero for a sign-changing gap function with $\emptyset(\mathbf{q})$ = $\pi$. As already mentioned, from the observed enhanced INS intensity at $\mathbf{Q_{AF}}$ **just** above the spin gap [38] it is concluded that the superconducting order parameter changes sign between two parts of the Fermi surface spanned by $\mathbf{Q_{AF}}$.

Different possible gap functions have been considered and the dynamical susceptibility calculated [48]. As a result, the $d_{x^2-y^2}$ symmetry of the superconducting gap has been found to best match the INS data of CeCu$_2$Si$_2$ [38]. Further support for *d*-wave symmetry of the superconducting order parameter comes from calculations solving the Eliashberg equations using the INS data, which revealed quite reasonable values for the superconducting $T_c$ [49].

Consequently, the superconducting order parameter of CeCu$_2$Si$_2$ was for a long time believed to correspond to a *d*-wave state with line nodes in the energy gap on the Fermi surface. The presence of line nodes was primarily inferred from measurements of the nuclear spin-lattice relaxation rate, 1/$T_1(T)$, which does not show a coherence peak below $T_c$ and exhibits a $\sim T^3$ power law behaviour down to about 0.12 K [50], see Fig. 2a [40].

Further support for line nodes in both *S* - and *A/S* - type CeCu$_2$Si$_2$ is provided by measurements of the thermal conductivity $\kappa(T,B)$ at elevated temperatures [51]. A strong decline of the thermal conductivity was observed below $T$ = 0.15 K for all samples studied and at all applied magnetic fields up to $B > B_{c2}$. This was ascribed to a decoupling of the charge carriers and the phonons due to deterioration, i.e., the development of a large electrical resistance of the contact of the interface between the heater and sample [52]. For this reason, the analysis of the $\kappa(T,B)$ data had to be confined to the temperature range $T$ > 0.15 K. In the following, we summarize the results obtained for an *S* - type CeCu$_2$Si$_2$ single crystal [51]. In the superconducting phase below $T$ = 0.5$T_c$ ≈ 0.3 K where the phonon contribution to the heat transport is negligibly small (a few %), one finds $\kappa(T)/T = a + bT$. This is in agreement with specific heat results in the same temperature range [53, 54]. The second term suggests the presence of line nodes in the gap function, while the first one indicates a broadening of the nodal regions on the Fermi surface caused by potential scattering, the latter being an *intrinsic* property of this compound due to the small (< 1%) site exchange between Cu and Si within the homogeneity range of CeCu$_2$Si$_2$ [55, 36]. As reflected by their large residual resistivities of $\rho_0$ ≈ 40 $\mu\Omega$cm [19], the Cu-Si site exchange is more pronounced for *S* - type samples due to their location away from the stoichiometric composition, whereas *A/S* - type samples with nearly stoichiometric composition exhibit $\rho_0$ values about four times smaller [19]. Because of the lack of access to temperatures below 0.15 K (where specific-heat measurements reveal a fully developed gap [53], see below), it is impossible to conclude from [51] whether the nature of the residual (as $T\to0$) value *a* is intrinsic or extrinsic. The *a*-value communicated in [51] may indeed be of extrinsic origin, i.e., caused by a fraction of about 10% of non-superconducting, antiferromagnetically ordered regions of the sample, phase separated from the majority superconducting part [38]. Very-low-temperature thermal-conductivity measurements, employing better heater-sample contacts, have been recently performed, which indeed revealed the lack of a residual term at zero temperature in the thermal conductivity [56].

The strong potential scattering inherent to $S$ – type $CeCu_2Si_2$ is most likely also the reason for the fact that *no* modulation of the thermal conductivity $\kappa$ as a function of the angle between the heat current and the direction of the magnetic field could be detected within the *ab*-plane deep inside the superconducting state of such $CeCu_2Si_2$ single crystals ($T$ = 0.15 K, $B$ = 0.2 T) [51]. On the other hand, a four-fold modulation of the upper critical field was observed from angle-resolved in-plane measurements of the electrical resistivity [57]. These results, which should be considered with some caution as the resistivity changes by an only very small amount (0.5 %), are consistent with a $d_{xy}$ pairing state [57]. In contrast, the already mentioned $d_{x^2-y^2}$ state inferred from the INS results [38] satisfies the requirement that the superconducting order parameter changes sign between the regions of the Fermi surface separated by $Q_{AF}$ [48]. In 2014, Kittaka et al. [53] performed state-of-the-art measurements of the specific heat down to temperatures as low as almost 30 mK on a high-quality $S$ - type $CeCu_2Si_2$ single crystal and observed an exponential temperature dependence, i.e., evidence for a node-less gap in the low-$T$ regime, $T \lesssim 60$ mK. Note that early point-contact studies already revealed a fully developed gap in $CeCu_2Si_2$ [58]. The temperature dependence of the specific heat coefficient $C(T)/T$, which early on was found to change from a $\sim T^2$ dependence slightly below $T_c$ to a power law dependence, $\sim T^\varepsilon$ with substantially larger exponent $\varepsilon \simeq 3$ at the lowest temperature, $T \simeq 50$ mK [59], can be fitted by an isotropic two-gap model [53]. Signatures of multi-gap superconductivity in $S$ - type $CeCu_2Si_2$ were also reported from subsequent scanning tunnelling spectroscopy (STS) measurements [60]. The results by Kittaka et al. [53] initiated a lively debate concerning the shape of the superconducting order parameter which we shall address below.

Obviously, the discrepancies between the pairing symmetries inferred from different probes illustrate that the superconducting order parameter of $CeCu_2Si_2$ is not well understood. For this reason it is necessary to reconcile in particular the observation of fully-gapped superconductivity with previous results suggesting a (one-band) nodal $d$ - wave pairing state. Clearly, a two-gap $s_{++}$ state of the type commonly applied to multi-band phonon-mediated superconductors does not change sign over the Fermi surface [44, 45] while for $CeCu_2Si_2$, with intra-band [17] rather than inter-band nesting, the proposed loop-nodal $s_{+-}$ state [61] does not appear to yield a node-less gap which changes sign over $Q_{AF}$.

The temperature dependence of the London penetration depth [$\Delta\lambda(T)$] is an important quantity for clarifying the superconducting gap structure, and several measurements have been recently reported on single crystals of $CeCu_2Si_2$ using a tunnel diode oscillator (TDO) based technique, which all also reveal full gapped behavior [19, 56, 62]. The low-temperature behaviour of $\Delta\lambda(T)$ for an $S$ - type crystal of $CeCu_2Si_2$ from Ref. 19 is displayed in Fig. 2d. The solid line shows a fit of the data by a fully gapped model up to $T_c/5$, with a gap magnitude 0.48 $k_BT_c$, consistent with the fully gapped behaviour found from the specific heat [53]. A fit to a power law dependence $\sim T^n$, with $n$ = 2.24, is also shown. When the data are fitted $\sim T^n$ from the base temperature up to a maximum temperature $T_{up}$, the fitting parameter $n$ increases with decreasing $T_{up}$ to values significantly larger than two, which is also consistent with fully gapped superconductivity [19].

To explain the seemingly conflicting experimental results, a 'd+d band-mixing' pairing state was proposed for CeCu$_2$Si$_2$ [19]. This state is analogous to an "$s\tau_3$" pairing state [43] advanced for the multi-orbital superconductivity of the iron chalcogenides. Expressed in the band basis, it is the sum of an *intra-band* pairing with $d_{x^2-y^2}$ symmetry and an *inter-band* pairing with $d_{xy}$ symmetry. It has an effective, finite gap in the quasiparticle-excitation spectrum because the two components appear as the sum in quadrature in the dispersion. Specifically, $\Delta(\phi)=[(\Delta_1\cos2\phi)^2 +(\Delta_2\sin2\phi)^2]^{0.5}$, where $\phi$ is the azimuthal angle. As concluded from Ref. 63, for CeCu$_2$Si$_2$ the projection of the AF ordering wavevector $\mathbf{Q_{AF}}$ onto the $k_x$- $k_y$ plane connects the corresponding projections of the parallel, flat parts of the dominating heavy-fermion pocket (warped cylinders) in the intra-band component (see Fig. 3), thereby generating a strongly enhanced spin spectral weight above a threshold (spin-gap) energy $\hbar\omega$ = 3.9 k$_B$T$_c$, substantially smaller than 2$\Delta_1$(0) $\simeq$ 5 k$_B$T$_c$.

As seen in Fig. 2e, the data of the normalized superfluid density $\rho_s = [\lambda(0)/\lambda(T)]^2$ are well fitted by this model using a cylindrical Fermi surface [64], with $\Delta_1(0)$ = 2.5 k$_B$T$_c$ and $\Delta_2(0)$ = 0.58 k$_B$T$_c$, the same results being calculated upon exchanging the two gap values. In addition, this model can explain our results of the temperature dependence of the specific heat of *S* - type samples. Fig. 2f displays 1 + $\delta C_{sc}/\gamma^*T$, where, following the method of [53], $\delta C_{sc}(T)$ is the difference between the electronic specific heats $C_{el}(T)$, measured at zero field and an applied (overcritical) field of 2 T. For $B$ = 2 T, $C_{el}(T)$ is obtained by subtracting the nuclear contribution from the measured data, yielding an effective Sommerfeld coefficient of $\gamma^*(T = 0)$ = 0.81 JK$^{-2}$mol$^{-1}$. As shown by the solid line, the data are well described using the *d+d* band-mixing pairing model with $\Delta_1(0)$ = 2.18k$_B$T$_c$ and $\Delta_2(0)$ = 0.56k$_B$T$_c$, taking into account a 9% non-superconducting volume fraction, in line with the results of the INS experiments discussed above [38]. No surprise, as was demonstrated in [19], this model can well account for the specific-heat data of Kittaka et al. [53] (and most likely also for the thermal conductivity results of [56]) with similar parameters. It should be noted that for the fit of this model to the data only two fit parameters are required, while for the two gap *s*-wave model used in Ref. 53 one needs three.

Figs. 2b and c display very recent results of the spin-lattice relaxation rate, $T_1(T_c)/T_1(T)$ vs. $T/T_c$, from $^{63}$Cu-NQR for an *S* - type CeCu$_2$Si$_2$ single crystal [65]. As clearly seen in Fig. 2b, below 0.2 K there is a significant deviation of this rate from the $T^3$ law established at elevated temperatures, cf. Fig. 2a. Both, a two-gap $s_\pm$ and the two-band *d+d* band-mixing model (with similar fit parameters as obtained from the penetration-depth data) describe the 1/$T_1(T)$ results well for temperatures $T$ < 0.8 $T_c$. However, the $s_\pm$ model [61, 66] cannot explain the spin-lattice relaxation rate data at elevated temperatures, where in the vicinity of $T_c$ it predicts a small, but finite Hebel Slichter peak which is not resolved from the experimental data (Fig. 2c) – in agreement with previous results [40, 50]. We note that in the case of the iron –based superconductors, it was proposed that quasiparticle damping can suppress the Hebel Slichter peak expected for the $s_\pm$ - state, and impurity-induced in-gap states can lead to an extended $T^3$ dependence of the spin-lattice relaxation rate [67]. However, whether such effects can account for the temperature dependence of 1/$T_1(T)$ for CeCu$_2$Si$_2$ across the whole temperature range has not been clarified, whereas these data can be well described by the *d+d* band-mixing model [19]. Because the *d+d* band-mixing pairing state can reconcile *all* the literature results, in particular the required sign change of the superconducting order parameter along $\mathbf{Q_{AF}}$ within the dominating heavy-fermion band

(see Fig. 3), it appears to be the prime candidate for the superconducting order parameter of CeCu$_2$Si$_2$.

For the heavy-fermion case we consider here, both the intra-band and inter-band *d*-wave pairing components will arise on the condition that the electronic states are within an energy range given by the RKKY interaction (more precisely, the spin excitation energy at the magnetic zone-boundary). This conclusion is based on calculations of Ref. 43, which showed the $s\tau_3$ pairing state to be energetically favoured in a multi-orbital model as long as the involved electronic bands are within an energy range of the AF exchange interaction *J*. In CeCu$_2$Si$_2$, the renormalized heavy-fermion states are located near the Fermi energy within the Kondo energy scale. Because the Kondo energy scale is comparable with the RKKY interaction, given that the system is close to an AF QCP, the afore-mentioned condition is readily satisfied and the *d+d* band-mixing pairing state represents a robust superconducting state. Interestingly, the effective gap of *d+d* band-mixing pairing is the same as that of a *d*+i*d* state, and the good fits to the temperature dependences of both the superfluid density (Fig. 2e) and specific heat (Fig. 2f) are also consistent with this latter state. There are, however, two issues with the *d*+i*d* state. (i) It breaks time-reversal symmetry. However, μSR measurements on *A/S* – type samples do not find evidence for broken time-reversal symmetry in the superconducting state [68], although further work along this line is under preparation. (ii) The *d*+i*d* state does not correspond to a single irreducible representation of the point group of CeCu$_2$Si$_2$ and would generally be expected to give rise to a second superconducting transition, which has not been experimentally observed.

We note in passing that the DFT+U calculations for the electronic structure, applied in the recent work of Y. Li et al [66], do not describe the Kondo renormalization effect; as a result, the calculated 4*f*-electronic states are not confined to the vicinity of the Fermi energy within the Kondo or RKKY energy scale. Thus, the approach of Li et al. cannot capture the *d+d* band-mixing pairing state. Li *et al.* also proposed an $s_{+-}$ pairing state for CeCu$_2$Si$_2$. There are two problems with this proposal. First, when an $s_{+-}$ pairing state is constructed as an irreducible representation of the point group associated with the lattice symmetry of CeCu$_2$Si$_2$, it is generically nodal, because the Fermi surface of CeCu$_2$Si$_2$ contains sheets which are quite extended in its Brillouin zone. This is to be contrasted with the case of the iron pnictides in which the Fermi surface comprises small pockets restricted to the center and boundaries of the Brillouin zone [69]. Second, as mentioned before, the wavevector $\boldsymbol{Q}_{AF}$ for the intensity peak in the INS spectra spans the Fermi wavevectors within the warped part of a **single** cylindrical sheet of the electron Fermi surface. Because the pairing function of Li et al. does not change sign within an electron Fermi surface (it does so only **between** the electron and hole Fermi surfaces), its sign is the same between the Fermi wavevectors connected by $\boldsymbol{Q}_{AF}$; thus, it will be unable to explain the intensity peak observed in the INS spectrum [38].

We close this subsection with two remarks concerning the complex response of superconductors with strong local Coulomb correlations to disorder. This is a pertinent issue, since two-gap superconductivity without a sign-changing order parameter has been proposed for CeCu$_2$Si$_2$, on the basis of experiments on electron-irradiated samples [56, 62]. These reported that the disorder induced by electron irradiation does not have a significant effect on $T_c$ [56], and does not induce low-energy bound states [62], which were both taken as evidence for a non-sign-changing superconducting order parameter.

Here, we first note that strong correlations make unconventional superconductors less sensitive to disorder associated with non-magnetic potential scattering than suggested by a straightforward generalization of the Abrikosov-Gor'kov theory for BCS superconductors with magnetic impurities [70]. In models with short-range spin-exchange interactions, such an effect has been explicitly shown [71, 72]. The $s\tau_3$ pairing state [43] arises from these type of exchange interactions and is expected to likewise be less sensitive to this kind of potential scattering. Given the strong correlations of the 4$f$-electrons in heavy fermion systems, the $d + d$ band-mixing pairing state proposed for CeCu$_2$Si$_2$ indeed appears to be robust against certain types of disorder, such as electron irradiation [56, 62] or, as already discussed above, variations of the residual resistivity [19] due to tiny changes of the ratio of Cu/Si site occupation inside the homogeneity range [36].

Second, in strongly correlated materials simple defects, e.g., atomic substitutions, strongly affect physical properties, cf. Zn substitution for Cu in the cuprate high-$T_c$ superconductors [73]. As already mentioned, in CeCu$_2$Si$_2$ very small amounts of non-magnetic as well as magnetic impurities suppress superconductivity. For example, in case of Rh, Pd and Mn substitution for Cu as well as Sc substitution for Ce the critical concentration at which $T_c \to 0$ was found to be as low as 1 at% [26]. On the other hand, Ce substitution was found to be extremely size-dependent: Upon increasing the size of the atomic radius from Sc to Y and further to La and Th, the critical concentration raises from 1 to 6 and further to 10 and 20 at%, respectively [26, 74]. This indicates that superconductivity is becoming increasingly suppressed by increasing chemical pressure [26].

The critical concentration for Ge partially replacing Si amounts to a value between 10 and 25 at% [75]. When reducing $T_c$ through 10 at% Ge-substitution by about 50% and applying additional hydrostatic pressure $p$ one, in fact, finds a narrow dome of superconductivity centred at a very low pressure value. This illustrates that the size effect which characterizes the substitution of Ce is indeed steric in origin. Interestingly, at high pressure a second dome of superconductivity with enhanced $T_c$ values and centred at about 5 GPa has been detected for CeCu$_2$(Si$_{0.9}$Ge$_{0.1}$)$_2$ [75] which is likely due to a nearby continuous valence transition at zero temperature [76]. This second dome has been also illustrated by applying hydrostatic pressure to pure CeCu$_2$Si$_2$, cf. Ref. 77. In the same paper, very similar behaviour at substantially larger pressure values is reported for the Ge-homologue, the pressure-induced superconductor CeCu$_2$Ge$_2$ [77].

### 3. YbRh$_2$Si$_2$: Heavy- fermion superconductivity near a zero-temperature Mott-type transition

***Unconventional quantum criticality.*** YbRh$_2$Si$_2$ is considered to be a canonical Kondo-lattice system, too. It exhibits a Kondo scale for the fully degenerate $j$ = 7/2 Hund's rule groundstate of Yb$^{3+}$ corresponding to 80 – 100 K [78]. According to an estimate of the magnetic entropy of the pure compound [18] and a thermopower study of Lu-substituted YbRh$_2$Si$_2$ [79] its single-ion Kondo temperature $T_K$ associated with the lowest-lying CF – derived Kramers doublet, which is well separated from the three excited ones, amounts to about 30 K. Very weak AF order with a staggered moment of $\simeq 0.002\mu_B$ [80] sets in at a much lower temperature, $T_N \approx 70$ mK [81]. Thermodynamic measurements prove that the Néel transition is of second order to the lowest temperature of these measurements ($\simeq 20$ mK) [82]. Therefore, it is natural to assume that $T_N$ can be smoothly suppressed by applying a small magnetic field ($B_N \approx 60$ mT || ab-plane, $\approx 660$ mT || c-axis) at which the QCP is accessed [83]. High-quality YbRh$_2$Si$_2$ single crystals with residual resistivities $\rho_0 \leq 1$ μΩcm are available [84]. Both the low-temperature paramagnetic phase ($B > B_N$) and the AF phase ($B < B_N$) behave as very heavy Fermi liquids [83] while a funnel-shaped quantum critical regime in the $B - T$ phase diagram with pronounced NFL phenomena is found to be centred at $B = B_N$, see Fig. 4a. Also shown (in black) in this figure is a sublinear crossover line $T^*(B)$ hitting, e.g., $\simeq 130$ mT at $T = 0.3$ K, which was constructed by the midpoints of thermally broadened jumps in the longitudinal magnetoresistance, MR [85]. The latter agree well with the midpoints of isothermally measured broadened jumps of the initial normal Hall coefficient, $R_H(B)$ [85, 86], as well as the positions of distinct anomalies in isothermal magnetostriction and magnetization data [82].

When the experiments are carried out to sufficiently low temperatures of the order of 50 mK, the $T^*(B)$ line is found to extrapolate (as $T \to 0$) to $B^*$ which is equal to the critical field of the AF order, $B_N(T_N \to 0) \simeq 60$ mT ($B \perp c$), within the experimental uncertainty [85]. For all measurements performed on samples of different quality, the jumps in both isothermal MR and $R_H(B)$ were found to remain finite, when extrapolated in the most reliable way to zero temperature [85]. At the same time these broadened jumps narrow substantially. The temperature dependence of their full width at half maximum (FWHM) can be best described as being *proportional* to temperature up to $T \approx 0.3$ K, where the jump in $R_H(B)$ begins to fade out (Fig. 4b), and up to $\approx 1$ K in MR [85]. In conclusion, these extrapolations of the data from $T > 0$ to $T = 0$ yield for *all* samples studied a finite and abrupt jump at the field-induced QCP [85, 86]. This jump in the Fermi surface volume indicated by the discontinuous change of $R_H(B)$ and MR as extrapolated to $T \to 0$ for $B^* \approx B_N$ was predicted for a Kondo-breakdown QCP [10, 11] at which the propagating Kondo singlets disintegrate, presumably under the influence of strong low-dimensional AF quantum critical fluctuations [10, 11]. In the Fermi liquid states on either side of the QCP Landau quasiparticles exist whose weight decreases continuously to zero at the QCP, where both Fermi surface volumes coexist [87]. The quantum critical fluctuations between the small and large Fermi surface volumes involve comparable weights of both even at $B = 0$ at a temperature as low as 0.5 K [87], see Fig. 4a. Since in general, the spectral weight of the small Fermi surface volume is quite small [89], and for YbRh$_2$Si$_2$ ($B = 0$) quantum critical fluctuations persist all the way down to $T_N$ [18], it appears to be almost impossible for angle-resolved photoelectron spectroscopy (ARPES) to observe a dominating small Fermi surface volume at $B = 0$ even at $T_N = 70$ mK. This is more than one order of magnitude lower than the current limit of ARPES measurements, $T \simeq 1$ K [90]. Therefore, the latter appear to be unapt to illustrate the abrupt change of the Fermi surface volume at the QCP of this material. While the authors of Ref. 90 were unable to monitor this most interesting change of the Fermi surface volume in YbRh$_2$Si$_2$ below 100 mK, they should have seen the necessary one associated with the crossover to the incoherent Kondo scattering regime at elevated temperatures, if they would have extended their measurements to $T > 90$ K. This change from large to small Fermi surface volume has recently been observed by ARPES measurements on CeCoIn$_5$ to occur at about 120 K [91] (See also Ref. 92 for CeRhIn$_5$).

The discontinuous change in Fermi surface volume discussed above is accompanied by a violation of the Wiedemann Franz (WF) law. The (in metal physics) fundamental WF law holds if the thermal transport in the zero-temperature limit is due to well-defined quasiparticles. In fact, the WF law has been shown to hold at sufficiently high fields ($B > 3B_N$) [87], and it is inferred to be valid as $T \to 0$ on the *whole* paramagnetic side ($B > B_N$) of YbRh$_2$Si$_2$, where a heavy Fermi liquid phase is caused to form due to the lattice Kondo effect [87]. As already mentioned, in the AF phase ($T < T_N$, $B < B_N$), a heavy Fermi liquid phase was also observed [18, 83]; consequently, the WF law must hold here in the zero-temperature limit as well. However, at $B = B_N$ the most reliable extrapolation of the finite-temperature results for $\rho(T)$ and the thermal conductivity $\kappa(T)$ to $T = 0$ show that the residual electronic thermal resistivity exceeds the residual electrical resistivity by about 10 % (here, the former is measured in the same units as the latter). Thus, as $T \to 0$ the Lorenz number $L = \rho\kappa/T$ deviates from Sommerfeld's constant $L_0 = (\pi^2/3)(k_B/e)^2$, and the WF law is violated exactly at the QCP [87].

This result was questioned by several groups [93 - 95], most recently by Taupin et al. [96, 97], all of whom report experimental data quite similar to those of [87]. The problem here is a bosonic term [$\kappa_m(T)$] to $\kappa(T)$ which, at low temperatures, masks the electronic part [$\kappa_{el}(T)$] but *must* vanish at $T = 0$ [87]. Low-temperature specific-heat measurements in the AF phase of YbRh$_2$Si$_2$ reveal a magnetic contribution ($C_m \sim T^3$) up to $T \simeq 50$ mK [18], whereby $\kappa_m(T)$ could be ascribed to acoustic AF magnons at $B < B_N$ and their overdamped counterparts ("paramagnons") at $B \gtrsim B_N$ [87]. In Refs. 93 and 94, this dominant paramagnon contribution to the low $- T$ $\kappa(T)$ at $B \simeq B_N$ was completely ignored, i.e., the measured $\kappa(T)$ values were erroneously identified with the electronic part $\kappa_{el}(T)$. According to Ref. 96, $\kappa_m(T)$ *increases* upon cooling to temperatures as low as 8 mK ($B = 0$) and 15 mK ($B = B_N$). On further cooling, $\kappa_m(T)$ has to pass over a maximum and assume its asymptotic ($T^3$) behaviour which in principle could be subtracted from the measured $\kappa(T)$ to determine the electronic thermal conductivity on the approach of the QCP. However, $\kappa(T)$ can hardly be measured at these extremely low temperatures. Therefore, it was proposed to extrapolate the data from a range of *elevated* temperatures, 0.1 – 1 K [87], where both the phonon contribution and $\kappa_m(T)$ can be neglected, while the electronic transport already follows the asymptotic quantum critical behaviour preceding the non-Fermi-liquid (NFL) ground state at $B = B_N$ [18], see Figs. 2a and b in Ref. 93 and Fig. 1 in Ref. 96. Here, both the electronic thermal resistivity and its electrical counterpart show a linear $T$ - dependence. In this temperature range one, therefore, observes a temperature-independent electronic Lorenz ratio which turns out to be 10% smaller than the WF value, i.e., $L_{el}(T)/L_0 = \rho(T)\kappa_{el}(T)/L_0 T \simeq 0.9$. The bosonic contribution $\kappa_m(T)$ is found to build upon this plateau below 70 - 100 mK, see the linear-linear plots of Fig. 11 in Ref. 95 and Fig. 2 in Ref. 96. Taupin *et al.* [96], however, consider $\kappa_m(T)$ to set in at a much lower temperature, below 30 mK. Thus, when extrapolating the Lorenz ratio from a temperature just above 30 mK, where the bosonic thermal conductivity already dominates, they accidentally arrive at an 'electronic' Lorenz ratio $\simeq 0.97$, i.e., a value very close to the WF value. They fit their data between 25 and 50 mK, but notice an increasing deviation of the fit from the data points upon warming. At $T = 90$ mK, this difference already amounts to $\simeq 5$%, which is of the same order as the deviation of $L(T \to 0)$ from $L_0$, reported in Ref. 87. This contradicts their conclusion that $\kappa_{el}(T)$ 'follows the WF law even at $B_N$' [96]. On the other hand, their results confirm those by Pfau

*et al.* [87] in that the WF law does hold far away from the QCP, i.e., in the heavy Fermi-liquid phase at $B > 3B_N$.

To minimize the ambiguities due to the extrapolation as a function of *T* as discussed before, an isothermal extrapolation from the high-field Fermi liquid regime into the low-field quantum critical regime was studied in Ref. 87 at temperatures where $\kappa_m(T)$ is negligible, i.e., from 100 to 400 mK: Here, $L(T)/L_0$ is typically smaller than one, indicating dominating inelastic scattering. Similar results had been reported in Ref. 93 where, however, the values extrapolated to *T* = 0 scatter around $L/L_0 = 1$ even at $B = B_N$ (Fig. 4 in [93]), because $\kappa_m(T)$ was erroneously considered to be part of and, therefore, was not subtracted from $\kappa_{el}(T)$.

Continuous isothermal $\rho(B)$ and $\kappa(B)$ measurements on YbRh$_2$Si$_2$ at a temperature as high as *T* = 490 mK [98] confirm the trends inferred from these *T* – scans taken at lower temperatures [87]: They reveal a minimum of $L(B)/L_0$ with increased width at a slightly higher field [98]. Thus, this minimum, which approaches ≃ 0.9 at 100 mK [87], appears to be a robust phenomenon. As already mentioned, right at $B = B_N$, the *residual* electronic thermal resistivity apparently exceeds the *residual* electrical resistivity by ≃ 10%, as nicely demonstrated by a distinct heating effect when adiabatically measuring the low-temperature electrical resistivity at low currents in the vicinity of $B_N$ [99], see Fig. 4c. In contrast to their main conclusion, the excellent data by Taupin *et al.* [96] confirm, rather than contradict, the violation of the WF law at the QCP of YbRh$_2$Si$_2$. This violation highlights additional scatterings, which naturally arise from the quantum critical fluctuations induced by the Kondo destruction in the local QCP description, i.e., fluctuations between a small and a large Fermi surface volume [10, 11]. The dynamical nature of the Kondo breakdown QCP as characterized by such *fermionic* quantum critical fluctuations is to be contrasted with the proposal, advocated in [100], of spin fluctuations bootstrapped by energy fluctuations, which assumes an incorrect singular form for the coupling between the electrons and energy fluctuations [101].

**Superconductivity at ultra-low temperatures.** No superconductivity was observed for YbRh$_2$Si$_2$ down to 10 mK, the lowest temperature reached in resistivity measurements utilizing a $^3$He-$^4$He dilution refrigerator [18]. This implies that the AF order below $T_N$ = 70 mK is *detrimental* to superconductivity. In order to search for superconductivity below 10 mK in this material, magnetic and calorimetric experiments down to *T* = 1 mK have been performed with the aid of a nuclear demagnetization cryostat providing a base temperature of 400 μK [20].

The field-cooled (fc) *dc*-magnetization of YbRh$_2$Si$_2$ registered at a field as low as 0.09 mT exhibits two phase-transition anomalies (Fig. 5a). The cusp at 70 mK displays the well-known Néel transition into what will be called in the following the 'primary electronic AF order'. The peak at ≃ 2 mK is ascribed to the transition into both the '*A* phase' (see below) and a superconducting phase. Zero-field cooled (zfc) *dc*-magnetization measurements indicate strong superconducting shielding below $T_c$ and partial shielding, i.e., the formation of small disconnected superconducting regions, below $T_B$ ≃ 10 mK. This is also reflected by the *ac*-susceptibility results obtained under nearly zero-field conditions (Fig. 5b). If the observed decrease of the (fc) *dc*-magnetization below $T_c$ measured in the smallest external field of 0.012 mT were ascribed completely to superconductivity and calibrated by the huge shielding signal found in the (zfc) *dc*-magnetization at the same field, a Meissner volume of, at most, 3% were obtained. Because of the concurring *A*-phase transition, the Meissner effect can contribute only partly to this decrease. However, a correspondingly small Meissner volume of order 1% only corresponds to what is commonly observed for bulk type-II superconductors with strong flux pinning, see, e.g., $CeCu_2Si_2$ [30]. Upon destroying the pinning centers of a type-II superconductor by powdering the sample and annealing the powder to release residual strains, one commonly observes an enormous increase of the Meissner volume, typically by much more than an order of magnitude [30], proving superconductivity to be a true volume effect. In conclusion, *bulk* superconductivity in $YbRh_2Si_2$ is all but unlikely; however this still requires verification by future (fc) *dc*-magnetization measurements on annealed powder samples.

The temperature scale $T_B \simeq 10$ mK is associated with an increase of the magnetization, when measured upon cooling (Fig. 5a). This is most likely caused by a decrease of the staggered magnetization of the primary *4f*-electronic AF order. The formation of disconnected superconducting regions below $T_B$, inferred from the (zfc) *dc*-magnetization results as mentioned before and also reflected by the *ac*-susceptibility data (Fig. 5b), may most naturally be related to this weakening of the primary order (which as mentioned before, appears to be detrimental to superconductivity) by the competing *A*-phase correlations. These are likely to nucleate around $T \simeq T_B$ at the randomly distributed Yb sites with finite nuclear spin values (see following paragraph). Upon cooling, these superconducting regions which are phase separated from the majority AF phase grow until at $T_A$, the long-range ordered *A*-phase develops, and superconductivity can form in the whole volume at $T_c \lesssim T_A$.

The data shown in Fig. 5c represent the molar specific heat of the Yb-derived nuclear spins as $\Delta C(T)/T$ measured at $B = 2.4$ mT; $\Delta C(T)$ was obtained by subtracting the nuclear quadrupole contribution (which is field dependent only in higher order) from the raw data at $B = 2.4$ mT [20]. Because of their weak hyperfine couplings, both the $^{103}$Rh and $^{29}$Si isotopes do not add to the specific heat in the temperature regime $T \geq 1$ mK. Here, only the contributions of $^{171}$Yb (nuclear spin $I = 1/2$) and $^{173}$Yb ($I = 5/2$) with natural abundances of about 15% each, are relevant – these nuclear Yb spins being randomly distributed. $\Delta C(T)/T$ displays a pronounced second-order phase transition at $T_A \simeq 2$ mK. The gigantic jump anomaly of $\simeq 1500$ JK$^{-2}$mol$^{-1}$ clearly indicates that this transition involves predominantly nuclear degrees of freedom. On the scale defined by the specific-heat results, it is clear that any *4f*-electronic and phonon contribution is completely negligible below about 10 mK [20]. Therefore, as no phase-transition anomaly is resolved at $B \simeq 60$ mT, this transition seen in the specific-heat results must indicate the onset of nuclear AF order. Interestingly, while the specific heat is probing only the nuclear degrees of freedom, the magnetic measurements presented in Figs. 5a and b display the response exclusively by the *4f* – electron spins: because of the small nuclear magnetic moment, a purely nuclear magnetic phase transition would not be resolved by either *dc*-magnetization or *ac*-susceptibility.

An alternative *à priori* possible scenario to interpret the specific-heat results involves a nuclear Kondo effect. Here, the specific-heat jump in Fig. 5c would display a superconducting transition, in accordance with the results of the magnetic measurements; in this case, the Cooper pairs had to be formed by super-heavy quasiparticles, conduction-electron spins entangled with nuclear (rather than 4*f*-electronic) spins. However, the associated nuclear Kondo temperature $T_{K,nucl}$ turns out to be many orders of magnitude smaller than $T_c \simeq$ 2 mK. Even if one assumes $T_{K,nucl}$ to be raised to a value as high as 25 mK, say, by some extremely strong renormalization of coupling parameters, the charge-carrier mass $m^*$ should still be enhanced over $m_e$, the mass of the bare electron, by a factor of $\simeq$ 400000. This lowest limit of $m^*/m_e$ expected by the action of a nuclear Kondo effect can be compared with $m^*/m_e$ estimated from the initial slope of the upper critical field for superconductivity at $T_c$, $B'_{c2}$ = - $(dB_{c2}/dT)_{Tc}$. The latter is obtained from both (fc) *dc*-magnetization measurements performed at very low fields (Fig. 5d) and the diamagnetic jumps in the (zfc) *M(T)* results (inset of Fig. 6). In either case, one finds $B'_{c2} \approx$ 25 T/K, very close to the value determined for, e.g., single-crystalline $CeCu_2Si_2$ [31]. This convincingly demonstrates a mass enhancement $m^*/m_e$ of the quasiparticles which is in the range of several 100 to 1000, typical for heavy fermions due to the ordinary Kondo entanglement between conduction-electron spins and 4*f* – electron spins, rather than nuclear spins.

Thus, a scenario based on a nuclear Kondo effect can be ruled out, and the transition at $T_A \simeq$ 2 mK (B = 2.4 mT) seen in the specific-heat data [20] has to be ascribed to a transition into AF order dominated by the Yb-derived nuclear spins which, at slightly lower temperatures, allows the formation of heavy-fermion superconductivity, as seen in the magnetic measurements [20]. According to Fig. 5c, the Yb-derived nuclear spin entropy saturates above 10 mK and becomes reduced by 26% upon cooling down to $T_A$ = 2 mK. This large nuclear spin entropy released in the paramagnetic range was ascribed to *A*-phase short-range correlations, which apparently compete with the primary electronic order and, as alluded to before, pave the way for superconductivity in disconnected regions below $T_B \approx$ 10 mK [20] (see Fig. 5b).

The existence of two different, closely spaced phase transitions is evident from the temperature dependence of the (fc) *dc*-magnetization obtained around 2 mK and at very low magnetic fields, as shown in Fig. 5d. It clearly exhibits a double peak structure where the sub-peak at elevated temperatures ($T_H$) can be related to the onset of the *A*-phase, while the minimum between the sub-peaks is considered a reasonable measure of the superconducting transition temperature $T_c$ [20]. As will become clear when discussing Fig. 6, the effective electronic *g*-factor of $YbRh_2Si_2$, $g_{eff} \leq$ 0.1, is much smaller than the in-plane 4*f*-electron *g*-factor ($g_{4f}$ = 3.5) of $YbRh_2Si_2$ [102]. On the other hand, $g_{eff}$ exceeds the value for a purely nuclear phase transition by a factor of order 50 [20]. Consequently, the *A* - phase though being dominated by the Yb-derived nuclear spins contains a tiny (2 – 3 %) 4*f* – electronic contribution, i.e., is a nuclear-electronic *hybrid* phase. In Fig. 5d, the onset of the hybrid *A* - phase at $T_A$ is identified with the change of slope in *M(T)* at the high-temperature side, due to the ordering of its small 4*f* – electronic component.

As already mentioned, one can identify the minimum position between the sub-peaks in Fig. 5d with $T_c$: In the Supplementary Material of Ref. 20 it is shown that the superconducting transition is of first order. Therefore, below $T_c$ any 4*f*-electronic order is likely to be expelled from the sample as is inferred from the strong increase in the measured magnetization, resembling the case of *A/S* – type $CeCu_2Si_2$ [37]. At slightly lower temperature, however, this increase is overcompensated by the decrease in *M(T)* due to the Meissner effect. The main results of these low-field (fc) *dc*-magnetization measurements are: (i) Below $B \simeq$ 4mT, there are two distinct subsequent phase transitions at $T_A > T_c$, $T_A \simeq 2.5$ mK and $T_c \simeq 2.0$ mK in the limit of zero field, while the (non-)existence of superconductivity at elevated fields has yet to be clarified by future resistivity measurements. (ii) The best linear fit through the $T_c(B)$ data yields an initial slope of $B_{c2}(T)$ at $T_c$ of − 25 T/K, typical of heavy-fermion superconductivity.

In Fig. 6, the *T* - *B* phase diagram is shown in a semi-logarithmic representation. The positions of the (fc) *dc*-magnetization peaks are displayed by the red symbols. For fields in excess of 4 mT, these peaks are not split and indicate the onset of the nuclear dominated AF *A* - phase. They are observed above the low-temperature limit of the experiments ($\simeq 1$ mK) up to 23 mT. Using a linear temperature scale, the peak positions can be extrapolated to $T = 0$, yielding a critical field value in the zero-temperature limit, $B_A$, between 25 and 55 mT. The ratio of the *A* - phase transition temperature at $B = 0$, $T_A$, to $B_A$ yields the afore-mentioned value of $g_{eff} \leq 0.1$. In Fig. 6, the blue hatching signifies the action of short - range *A* - phase correlations up to temperatures as high as $T_B$, as inferred from the substantial values of the Yb-derived nuclear spin entropy [20]. These short - range *A* - phase correlations are presumably competing strongly with the primary 4*f*-electronic AF order and, this way, cause the formation of disconnected superconducting regions.

The above qualitative conclusions drawn from the experimental results are most convincingly verified by a three-component Ginzburg-Landau (GL) theory [20], the minimal GL theory apt to explain the occurrence of two subsequent magnetic phase transitions (at $\simeq$ 70 and $\simeq 2$ mK). At the level of the free energy functional, short - range correlations are not taken into account. The theory relies on a very natural assumption, i.e., that the 4*f*-electronic spin susceptibility is *anisotropic*: The latter indeed exhibits a maximum not only at the AF ordering wave vector, ***q*** = ***Q*_{AF}**, of the primary electronic Néel phase (order parameter $\Phi_{AF}$) but also at ***q*** = 0, as illustrated by strong ferromagnetic quantum critical fluctuations [103]. It is, therefore, straightforward to assume that there exists another relative maximum in the spin susceptibility at a finite wave vector ***q*** = ***Q*_1**, different from ***Q*_{AF}**. In this case, the 4*f*-electron spins will be subject to an RKKY interaction along ***Q*_1**, mediated at these low temperatures by the strongly renormalized charge-carrier system and establishing an electronic order parameter $\Phi_J$. The same RKKY polarizations will also be experienced by the randomly distributed Yb-derived nuclear spins residing on the same lattice sites as 4*f* electrons and give rise to the formation of a nuclear order parameter $\Phi_I$. The coupling between two order parameters with the same wave vector is bilinear in GL theory, the coupling parameter here being determined by the very large on-site Yb hyperfine coupling parameter $\lambda$ (corresponding to $\simeq 25$ mK [20]). It yields a phase transition temperature for the hybrid *A*-phase $T_{hyb} \approx 1$ mK, which agrees with the experimental value $T_A \geq 2$ mK (Fig. 5) within a factor of 2.

Interestingly, this value for $T_{hyb}$ is obtained exclusively from experimental input parameters ($\lambda$, $g_{4f}$), the only unknown parameter being the 4*f* - electronic spin susceptibility at ***Q*_1**, which was therefore assumed to be equal to the bulk susceptibility. The wide parameter space of the GL theory allows one to choose the nuclear order parameter $\Phi_I$ to be much larger than the electronic $\Phi_J$, in accordance with the huge values of the measured specific heat at very low temperatures (Fig. 5c). Further on, the hybrid order can compete with the primary order, leading to a strong reduction of the staggered magnetization $m_{AF}$ at $T_{hyb}$ (Fig.7). In the real system the decline of $m_{AF}$ sets in already near $T_B \simeq 10$ mK due to the competing short - range $A$ - phase correlations. In the zero-temperature limit, $m_{AF}$ eventually even vanishes, establishing the underlying AF QCP, so that the concomitant quantum critical fluctuations may substantially contribute to the Cooper pairing.

## 4. Perspective

Transport [79, 104, 105] and STS [78] measurements on stoichiometric, clean heavy-fermion compounds of $Ce^{3+}$ and $Yb^{3+}$ commonly reveal lattice Kondo coherence to set in at the single-ion Kondo temperature $T_K$ associated with the CF - derived lowest-lying Kramers doublet [106]. The low-temperature state of such materials can be manifold, e.g., magnetically ordered, non-magnetic heavy Fermi liquid, non-Fermi liquid and superconducting [2, 4, 5]. The heavy Fermi liquid phase refers to a large Fermi surface volume to which delocalized 4$f$ states contribute as a consequence of the lattice Kondo effect [107]. On the other hand, well above $T_K$ these materials behave like dilute Kondo alloys with incoherent transport properties. Here, the Fermi surface volume is assumed to be small, due to the ordinary conduction electrons. For $CeCoIn_5$, a corresponding, thermally driven change from large to small Fermi surface volume has been inferred from STS [108] and, as already mentioned, also from ARPES [91] results. It is fair to say, however, that owing to their limited temperature range and energy resolution, current ARPES experiments [90] are unapt to distinguish between a conventional 3D-SDW QCP and its unconventional Kondo-destroying counterpart [88]. In the present paper, a variety of more suited experimental techniques were selected to unravel the low-temperature normal-state properties out of which superconductivity develops in both $CeCu_2Si_2$ and $YbRh_2Si_2$, the prototype systems for either of the two very different QCP scenarios as yet investigated for heavy-fermion metals.

In the case of $CeCu_2Si_2$ ($T_K \simeq 20$K), the universal $\omega/T^{3/2}$ scaling of its 4$f$-electron spin susceptibility [39] (Fig. 1b) as well as the afore-mentioned $T$ - dependences of both the temperature coefficient of the electronic specific heat and the electrical resistivity [35] yield convincing evidence for an *itinerant* 3D-SDW QCP at zero magnetic field and close to the stoichiometric composition [109]. The multi-band superconducting phase which develops in the vicinity of this QCP has $d$ - wave symmetry, with a fully developed gap at very low temperatures. This highlights a novel type of unconventional superconductor. The important conclusion that the gain in exchange energy below the superconducting critical temperature in $CeCu_2Si_2$ exceeds the superconducting condensation energy by a factor of order 20 has been ascribed to a correspondingly huge increase of kinetic energy, i.e., a break-up of Kondo singlets, necessary to form the superconducting state [38]. Moreover, the involvement of Mott - like physics is evident from the fact that $T^*$, the Fermi-surface crossover temperature and finite - $T$ signature of a Kondo-destroying QCP (see Sec. 3), is non-zero but quite small at a heavy-fermion SDW QCP [5]. Naturally, $T^*$ is much smaller than $T_K$ which implies that only long-lived spin fluctuations (with $\omega < k_B T^*/\hbar$) can be of the SDW type. On the other hand, the high-frequency part of the INS spectrum, which in $CeCu_2Si_2$ ranges up to $\simeq 2$ meV corresponding to $k_B T_K$ [38], represents Mott - like fluctuations of local 4$f$-electron spins associated with the break-up of the Kondo effect which is expected to occur not too far away in the phase diagram. This seems to be a generic property of any heavy-fermion metal displaying superconductivity near a 'conventional' 3D-SDW QCP.

For YbRh$_2$Si$_2$ ($T_K \simeq$ 30 K) an unconventional, i.e., *local* QCP [$T^*(B) \rightarrow$ 0] [10, 11] is inferred from an abrupt change of the Fermi surface volume at the critical field $B_N$ where $T_N$ smoothly vanishes. This is best illustrated by extrapolating isothermal magnetotransport results to zero temperature [85]. There are additional puzzling observations on YbRh$_2$Si$_2$ which cannot be explained within the itinerant quantum critical framework [6 - 8]. As striking examples, among others, we mention disparate temperature dependencies of thermodynamic and transport properties upon the approach of the QCP [18] and an unusually large critical exponent of the specific heat at the classical Néel transition ($T_N$ = 70 mK, $B$ = 0) in YbRh$_2$Si$_2$ which obviously is sufficiently close to its unconventional QCP ($T$ = 0, $B_N$ = 60 mT$\perp c$) [110].

The apparent violation of the WF law exactly at the field-induced QCP of YbRh$_2$Si$_2$ provides insight into the dynamics underlying this Mott - type transition. Here, at $T$ = 0, *fermionic* quantum critical fluctuations between small and large Fermi surface volume are strong scatterers for the electronic heat carriers and cause an increase of the residual electronic thermal resistivity compared to its electrical counterpart, leading to a Lorenz ratio $L(T\rightarrow 0)/L_0 \simeq$ 0.9 [87]. These novel kind of quantum critical fluctuations may be considered the *replica of additional inelastic scatterings* which give rise to a distinct minimum in the isothermal field dependence of $L(B)/L_0$ at low but finite temperatures [87, 98]. The latter processes are at the origin of the unusual heating effect observed on approaching the QCP [99], see Fig. 4c. One can compare YbRh$_2$Si$_2$, where the violation of the WF law is ascribed to a *local* QCP [87, 99], with other HF metals. For ß -YbAlB$_4$, this law seemingly holds [111] which, together with the [$\rho(T)$-$\rho_0$)] $\sim T^{3/2}$ dependence above $T_c$ = 80 mK [112], questions a local nature of its QCP. On the other hand, the observation of $L(T\rightarrow 0)/L_0 \approx$ 0.9 at a field-induced quantum bicritical point ($\approx$ 4.5 T) in YbAgGe [113] demonstrates that the violation of the WF law is not confined to a single material.

From the fact that no superconductivity had been detected in YbRh$_2$Si$_2$ at temperatures as low as 10 mK [18], one might have argued that the quantum critical fluctuations in this compound do not promote superconductivity. CeCu$_{5.9}$Au$_{0.1}$ also does not show superconductivity. In this case the lack of superconductivity, however, may be ascribed to the high level of alloying-induced disorder which depresses $T_c$, at least to below 20 mK [114]. From the recent discovery of superconductivity in YbRh$_2$Si$_2$ below $T_c$ = 2 mK [20] several lessons can be learned. (i) The emergence of superconductivity is governed by the system of the Yb-derived nuclear spins: rather than giving rise to a nuclear Kondo effect and super-heavy Cooper pairs, the nuclear dominated *A* - phase competes and in the zero-temperature limit eventually fully suppresses the primary 4$f$-electronic AF order. This competition between the *A* - phase and the primary Néel order illustrates a new way to reach the QCP in YbRh$_2$Si$_2$. Traditionally, this QCP is approached by applying a pair-breaking magnetic field, unapt to generate superconductivity as is evident from Fig. 6. (ii) Since $T_c$ is limited by the nuclear dominated ordering temperature $T_A \gtrsim$ 2 mK, the 'intrinsic' $T_c$ of YbRh$_2$Si$_2$ (which one would observe at $B$ = 0 in the putative paramagnetic state, with the 4$f$-electronic primary order nearby in the phase diagram) is very likely substantially higher. (iii) Superconductivity at an AF QCP, regardless of its microscopic nature, i.e., being either of the conventional SDW or the unconventional Mott type, appears to be a general phenomenon.

Hopefully, more Yb-based heavy-fermion superconductors will be discovered in the future to add to YbRh$_2$Si$_2$ [20] and $\beta$-YbAlB$_4$ [112].

**Acknowledgments:** The authors acknowledge stimulating discussions with Piers Coleman, Philipp Gegenwart, Silke Paschen and Doug Scalapino. M. Smidman, H. Q. Yuan, S. Kirchner, and F. Steglich acknowledge support from the National Key R&D Program of China (Nos. 2016YFA0300202 and 2017YFA0303100). M. Smidman, H. Q. Yuan and F. Steglich acknowledge support from the National Natural Science Foundation of China (Nos. U1632275 and 11474251), and the Science Challenge Project of China (No. TZ2016004). S. Kirchner acknowledges support by the National Natural Science Foundation of China (No. 11474250 and No. 11774307). Work done at the MPI CPfS was partially supported by the German Research Foundation through the DFG Research Unit 960 "Quantum Phase Transitions". Work at Rice University was in part supported by the NSF Grant No. DMR-1611392 and the Robert A. Welch Foundation Grant No. C-1411.

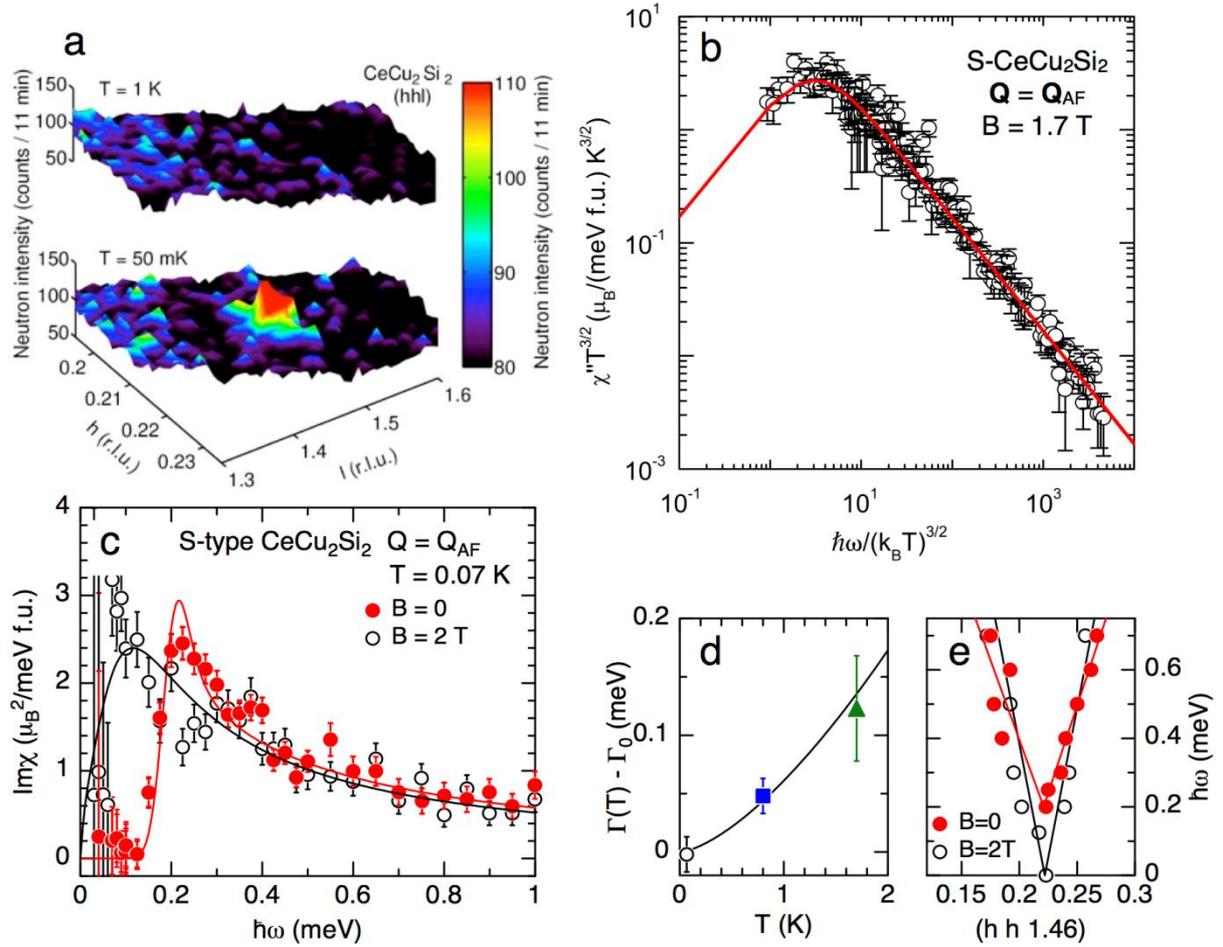

Fig. 1. Neutron scattering on single-crystalline $CeCu_2Si_2$: (a) Colour-coded intensity maps of the reciprocal ($hhl$) scattering plane in $A$-type $CeCu_2Si_2$ below and above $T_N \approx 0.8$ K showing the AF superstructure peak at $Q_{AF}$ = (0.215 0.215 1.47) at low $T$ = 0.05 K. Figure reproduced from O. Stockert *et al.*, Phys. Rev. Lett. 92 (2004), p. 136401 [17]. Copyright 2004 by the American Physical Society. (b) Scaling plot of the normal-state quasielastic magnetic response in $S$-type $CeCu_2Si_2$ at $Q_{AF}$ and $B = B_{c2} \simeq 1.7$ T. Displayed is the imaginary part of the dynamical susceptibility $\chi''$ as $\chi''T^{3/2}$ versus $\hbar\omega/T^{3/2}$. Figure reproduced from J. Arndt *et al.*, Phys. Rev. Lett. 106 (2011), p. 246401 [39]. Copyright 2011 by the American Physical Society. (c) Magnetic response of $S$-type $CeCu_2Si_2$ at $Q_{AF}$ and $T$ = 0.07 K in the superconducting ($B$ = 0) and the normal state ($B$ = 2 T). (d) Relaxation rate $\Gamma$ of the normal-state quasielastic magnetic response in $S$-type $CeCu_2Si_2$ at $Q_{AF}$ as a function of temperature indicating a $T^{3/2}$ dependence of $\Gamma$ (solid line). Plotted is the relaxation rate after subtraction of a residual $\Gamma_0 \approx 0.11$ meV, i.e., $\Gamma(T) - \Gamma_0$. The small, but finite $\Gamma_0$ indicates that the $S$-type crystal is located slightly on the paramagnetic side of the QCP. (c) and (d) are reproduced from O. Stockert *et al.*, Nature Phys. 7 (2011), p 119-124 [38]. Copyright 2011 by Springer Nature. (e) Dispersion of the magnetic excitations around $Q_{AF}$ at $T$ = 0.06 K in the superconducting and normal state of $S$-type $CeCu_2Si_2$ [39].

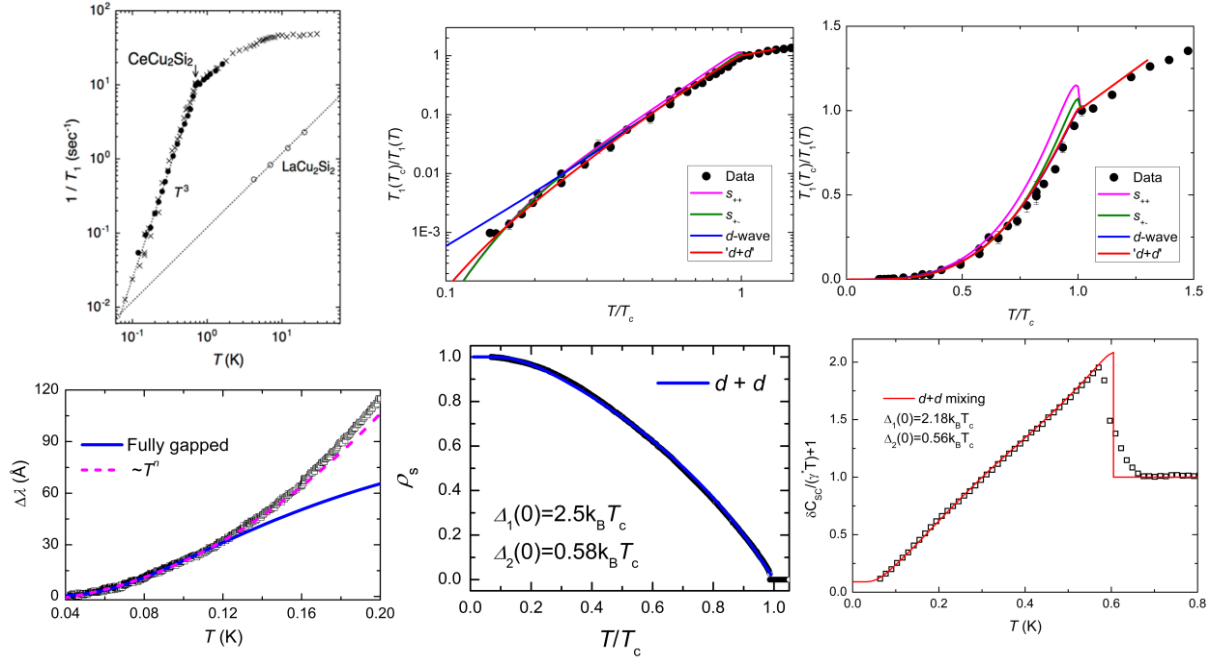

Fig. 2. Superconductivity in S - type $CeCu_2Si_2$: Temperature dependence of (a - c) the spin – lattice relaxation rate determined by Cu – NQR; a) is from Ref. 40, and (b, c) are from Ref. 65. (d) London penetration depth change $\Delta\lambda(T)$ [19] (solid and dashed lines show fits by a fully-gapped model and a $\sim T^n$ dependence, respectively), (e) normalized superfluid density $\rho_s(T)$ [19], and (f) specific-heat coefficient [54] as $1 + \delta C_{sc}/(\gamma^* T)$ vs. T, where the solid lines in (e) and (f) show fits to the d+d band-mixing pairing model [19].

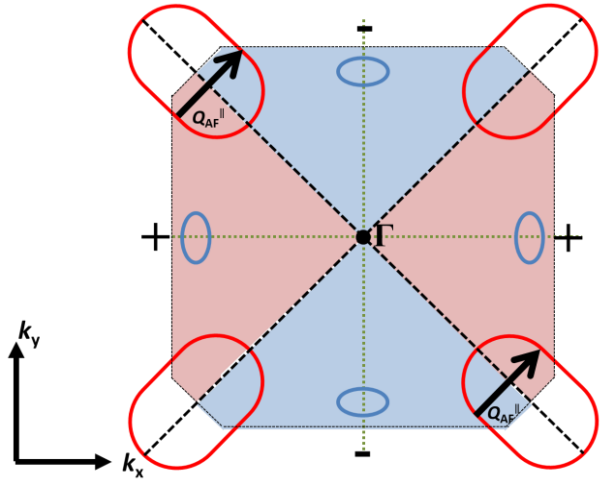

Fig. 3. Illustration of the renormalized (heavy) Fermi surface of $CeCu_2Si_2$. Warped, nesting parts of the dominating cylindrical Fermi surface (red) at particular values of $k_z$ as well as additional small pockets (blue), projected onto the $k_x - k_y$ wave-vector plane [63]. The component $Q_{AF}^{\parallel}$ of the AF ordering wavevector $\mathbf{Q}_{AF}$ projected onto the same wave-vector plane connects the parts of the heavy Fermi surface with a sign change in the intra-band pairing component. The two shaded regions illustrate the sign of this intra-band component [19].

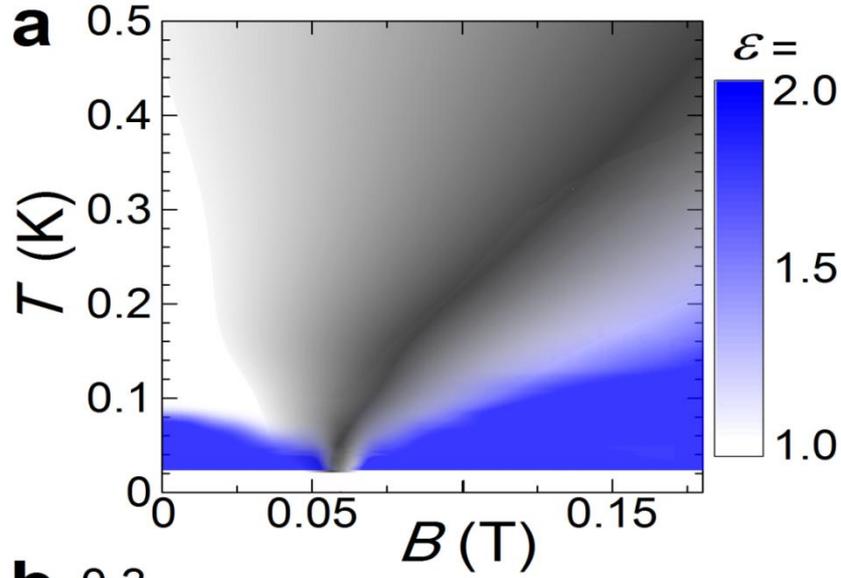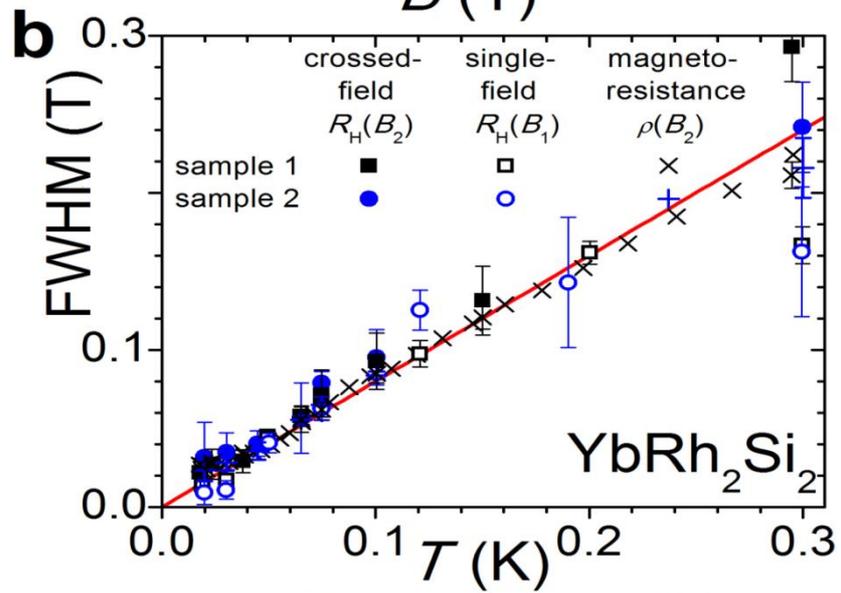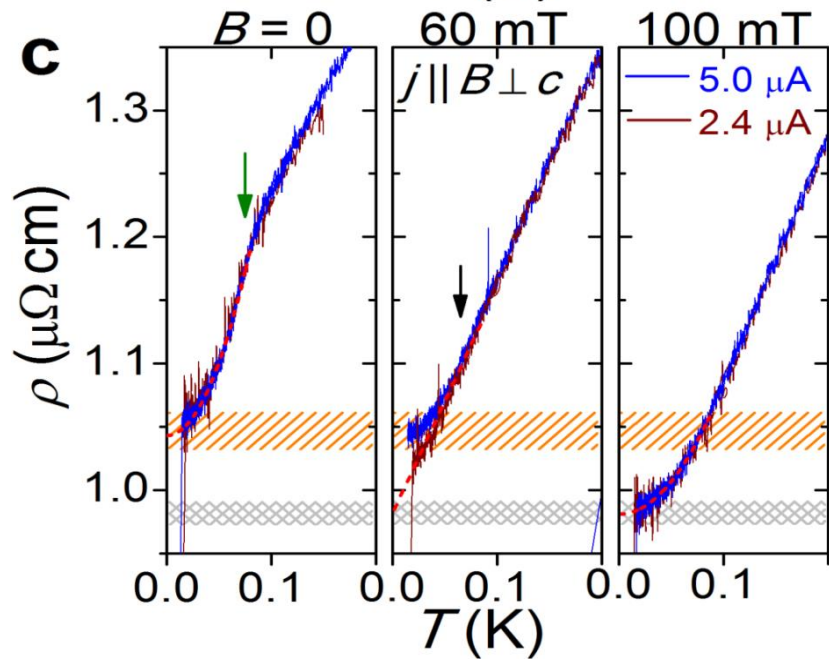

Fig. 4. Unconventional quantum criticality in YbRh$_2$Si$_2$. (a) Temperature ($T$) – magnetic field ($B$) phase diagram of a single crystal with residual resistivity $\rho_0 \simeq 1$ μΩcm, $B\perp c$. Colour code describes $\varepsilon$, the exponent in $\Delta\rho(T) = [\rho(T) - \rho_0] \sim T^\varepsilon$. Both the antiferromagnetically ordered ($T < T_N = 70$ mK, $B < B_N \approx 60$ mT, $\mu_{ord} \approx 0.002\mu_B$ [80]) and the low – $T$ paramagnetic phase ($B > B_N$) behave as heavy Fermi liquids (FLs) (blue: $\varepsilon = 2$), while the quantum critical regime shows non – Fermi –liquid (NFL) behaviour (white, gray, black: $\varepsilon \simeq 1$). The gray scale is a measure of the slope of the isothermal longitudinal magnetoresistivity (MR) associated with a thermally broadened crossover of the Fermi surface volume between small (low fields) and large (elevated fields) [85]. The dark line where this slope is largest denotes the crossover line $T^*(B)$ which is also obtained from the initial normal Hall-coefficient [86, 85] and thermodynamic measurements [82]. Note that the crossover between the NFL and FL regimes at $B > B_N$ is rather broad whereas the second-order phase transition at $T_N(B)$ is manifested by a rapid change in $\varepsilon$ between 1 and 2. Figure reproduced from Custers et al., Nature 424 (2003), p. 524-527 [18]. Copyright 2003 by Springer Nature. (b) Full width at half maximum (FWHM) of the crossover in isothermal magneto-transport results, i.e., normal Hall coefficient determined in both the crossed-field and single-field configuration as well as longitudinal MR for two single crystals. The solid line illustrates that the FWHM is proportional to $T$, with the data being well described up to 1 K. Figure reproduced from S. Friedemann et al., Proc. Natl. Acad. Sci. USA 107 (2010), p.14547-14551 [85]. (c) Temperature dependence of the electrical resistivity below $T$ = 0.1 K with two different low excitation currents (blue and brown traces, respectively) at several magnetic fields ($\perp$ c-axis): $B$ = 0, $B_N$ = 60 mT, $B$ = 100 mT. Where blue and brown traces overlap, the brown ones are extrapolated (in red) to $T$ = 0 assuming a $[\rho(T)-\rho_0] = AT^2$ dependence. The splitting between brown and blue traces at the critical field $B_N$, below $T$ = 70 mK (marked by black arrow) illustrates a heating of the sample under the applied larger current. This illustrates the violation of the Wiedemann Franz law at the QCP in YbRh$_2$Si$_2$ [87]. Note that this heating effect is absent away from $B_N$. The onset of AF order at $T_N$ = 70 mK is indicated by green arrow. Horizontal grey hatching displays an almost field-independent residual ($T \to 0$) resistivity in the paramagnetic regime. The weak field dependence of $\rho_0$ inside the AF phase is masked by the width of the orange hatching. The difference between the hatched horizontal regions indicates an abrupt decrease in $\rho_0(B)$ upon increasing the field through $B_N$. This illustrates a corresponding abrupt increase in the charge-carrier concentration at $B = B_N$ and $T$ = 0, i.e., at the field-induced QCP [99].

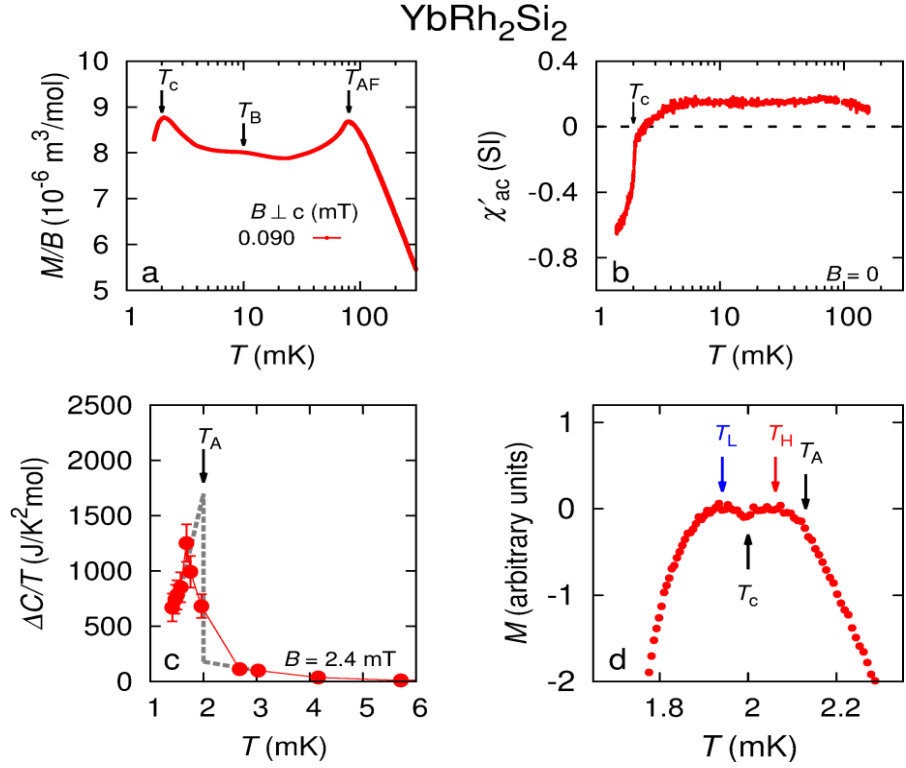

Fig. 5. (a) Field-cooled (fc) *dc*-magnetization curve of YbRh$_2$Si$_2$ taken at $B$ = 0.09 mT applied within the basal plane. There are three main features: the AF phase transition at $T_N = T_{AF}$ = 70 mK, a shoulder in magnetization at $T_B \simeq$ 10 mK and a pronounced peak at $T_c$ = 2 mK. (b) In-phase signal $\chi'_{ac}(T)$ of the *ac*-susceptibility measured after having compensated the earth – magnetic field. The features seen at $T_{AF}$, $T_B$ and $T_c$ in the *dc*-magnetization are detected by the *ac*-susceptibility, too. The large negative values of $\chi'_{ac}(T)$ indicate pronounced superconducting shielding below $T_c$. Both panels are reproduced from E. Schuberth et al., Science 351 (2016), p.485-488 [20]. Copyright 2016 AAAS. (c) $\Delta C(T)/T$ obtained after subtracting the nuclear quadrupolar contribution calculated for $B$ = 0 from the data taken at $B$ = 2.4 mT. A peak in $\Delta C(T)/T$ occurs at 1.7 mK. Assuming the transition to be of second order, an equal-area construction (see dashed line) yields a nuclear phase transition temperature $T_A \simeq$ 2mK. This almost coincides with the peak position found in the (fc) *dc*-magnetization. The associated jump of $\Delta C(T)/T$ is about 1500 J/K$^2$mol$^{-1}$. (d) (fc) *dc*-magnetization taken at a very low field of $B$ = 0.09 mT plotted as a function of $T$ in the vicinity of $T_c$ = 2 mK; $T_A$ is determined by the temperature where $M(T)$ changes slope at the high-field side, and $T_c$, determined by shielding experiments, i.e., of the *ac*-susceptibility and zero-field-cooled (zfc) *dc*-magnetization, coincides with the position of the minimum. The increase of $M(T)$ upon cooling at the high-field side reflects the decrease of the staggered magnetization of the primary order (due to the competing short - range A - phase correlations) around and below $T_B \approx$ 10 mK. The broad maximum at $T_H$ (red arrow) results from an overcompensation of this increase in $M(T)$ by the reduction of the magnetization upon the onset of the small (2 - 3%) 4$f$-electronic component of the hybrid order at $T_A$. Just at $T_c$, another increase of the (fc) *dc*-magnetization points to an additional reduction of the staggered magnetization, related to the fact that the superconducting phase transition is of first order. Finally at $T_L$ (blue arrow), slightly below $T_c$, $M(T)$ decreases again due to the expulsion of magnetic flux (Meissner effect) (from [20]).

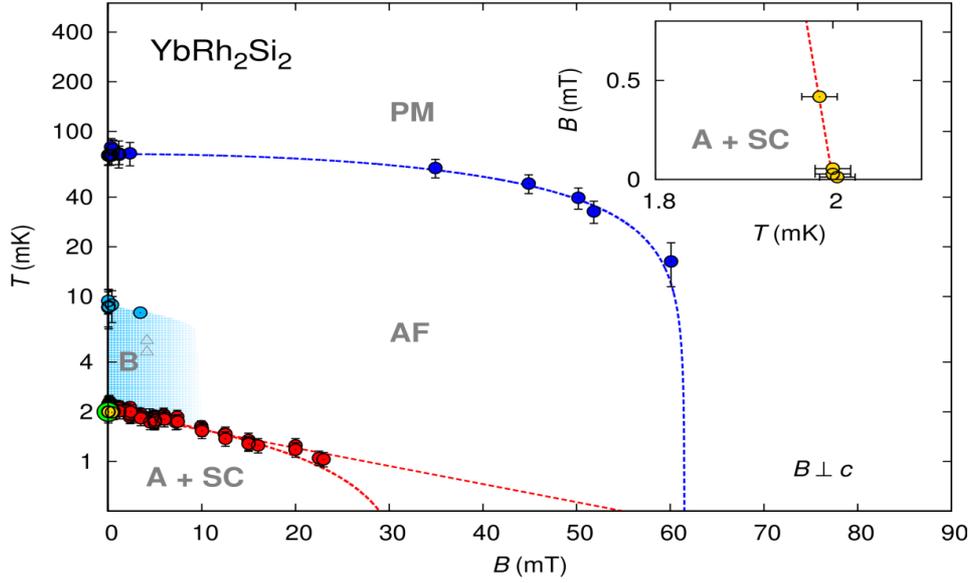

Fig. 6. Generic $T - B$ phase diagram of YbRh$_2$Si$_2$. It was obtained from *dc*-magnetization and *ac*-susceptibility measurements at several magnetic fields. 'AF' indicates the electronic primary AF order ($T_N$ = 70 mK), 'PM' indicates the paramagnetic state. The hatched light-blue area indicates the onset of short - range *A* - phase correlations which give rise to a reduction of the staggered magnetization and a splitting of the (zfc) and (fc) *dc*-magnetization curves, i.e., the beginning of shielding due to disconnected superconducting regions, see text. The two data points (gray triangles) determined via field sweeps of the *dc*-magnetization between 3.6 mK and 6.0 mK are most likely not related to those *A* - phase correlations [20]. 'A+SC' indicates concurring nuclear AF order and superconductivity, at least at fields below $\simeq$ 4 mT. At higher fields, no splitting is resolved in the low-temperature peak of the (fc) *dc*-magnetization, and its position is considered to be equal to $T_A$, marking the onset of long - range *A* - phase order. The two red dashed lines mark the range within which the boundary line of the *A* phase may terminate as $T \rightarrow 0$. The green circle indicates the superconducting transition temperature seen in the *ac*-susceptibility at $B$ = 0 while the yellow circles (partially covered by the green point) result from the shielding signals in the (zfc) *dc*-magnetization. In the inset, the positions of these shielding transitions are shown on an enlarged scale. The big slope of the superconducting phase boundary $T_c$ vs. $B$ at low fields, $B'_{c2}$ = (- $dB_{c2}/dT$)$_{Tc}$ $\simeq$ 25 T/K is typical for heavy-fermion superconductors [31]. The figure is reproduced from E. Schuberth *et al.*, Science 351 (2016), p.485-488 [20]. Copyright 2016 AAAS.

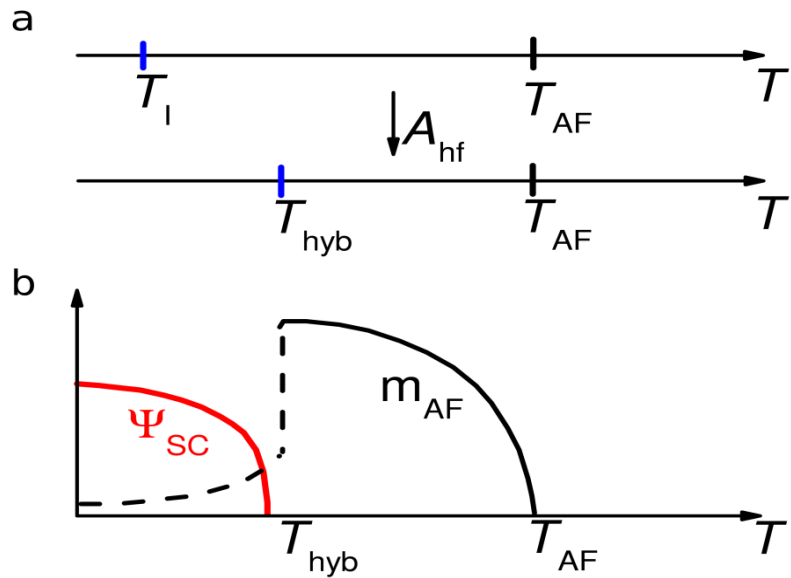

Fig. 7. (a) Sketch of the two phase transitions associated with $4f$ - electronic and nuclear spin orders. Top line: without a hyperfine coupling ($A_{hf}$), the electronic and nuclear spins are ordered at $T_{AF}$ and $T_I$, respectively. Bottom line: with a hyperfine coupling, $T_{AF}$ is not affected, but a hybrid nuclear and $4f$ - electronic spin order is induced at $T_{hyb} \gg T_I$. (b) Temperature evolution of the primary electronic spin order parameter $m_{AF}$ and the superconducting order parameter $\psi_{SC}$. $\psi_{SC}$ develops when $m_{AF}$ is suppressed by the formation of the hybrid nuclear – $4f$ electronic spin order right below $T_{hyb}$. The figure is reproduced from E. Schuberth *et al.*, Science 351 (2016), p.485-488 [20]. Copyright 2016 AAAS.